\documentclass[a4paper]{jpconf}
\usepackage{graphicx}

\begin{document}
\title{Hard X-ray morphology of the X1.3 April 25, 2014 partially
  occulted limb solar flare}

\author{Frederic Effenberger, F\'atima Rubio da Costa and Vah\'e Petrosian}

\address{Department of Physics and KIPAC, Stanford University,
  Stanford, CA 94305, USA}

\ead{feffen@stanford.edu}

\begin{abstract}
  At hard X-ray energies, the bright footpoint emission from solar
  flare loops often prevents a detailed analysis of the weaker
  loop-top source morphology due to the limited dynamic range
  available for X-ray imaging. Here, we study the X1.3 April 25, 2014
  flare with the \emph{Reuven Ramaty High-Energy Solar Spectroscopic
    Imager} (\emph{RHESSI}). This partially occulted limb flare allows
  the analysis of the loop-top emission in isolation. We present
  results on the flare light curve at different energies, the source
  morphology from X-ray imaging and a detailed spectral analysis of
  the different source components by imaging spectroscopy. The
  loop-top source, a likely site of particle acceleration, shows a
  clear composition of different emission components. The results
  indicate the opportunities that detailed imaging of hard X-rays can
  provide to learn about particle acceleration, transport and heating
  processes in solar flares.
\end{abstract}

\section{Introduction}
Hard X-rays provide one of the most direct diagnostics for energetic
particle processes in solar flares \cite{Fletcher-etal-2011}. Studies
in the past have shown that in some flares at least two distinct types
of sources, namely from the coronal solar flare loop-top and
chromospheric footpoints, can be distinguished
\cite{Frost-Dennis-1971,Roy-Datlowe-1975,McKenzie-1975,
  Hudson-1978,Masuda-etal-1994,Petrosian-etal-2002,Krucker-etal-2007,
  Simoes-Kontar-2013}. It is suggested by theory and observations that
the coronal region at the loop-top is the main acceleration site for
electrons (see, e.g. the review \cite{Petrosian-2012}). However, due
to the limited dynamical range of \emph{RHESSI} \cite{Lin-etal-2002},
it is often hard to clearly distinguish a coronal source, once strong
footpoint emission is present. Partially occulted flares, in which
the footpoints are a few degrees behind the solar limb, offer the
opportunity to observe the coronal sources in isolation. Krucker \&
Lin \cite{Krucker-Lin-2008} performed a statistical study of partially
occulted flares during solar cycle 23. They concluded that thin-target
radiation of electrons is a plausible scenario for the coronal
emission. Some of the analyzed flares also show significant structure,
indicating multiple coronal source regions. Kappa-distribution models
\cite{Kasparova-Karlicky-2009,Oka-etal-2013,Bian-etal-2014} have been
discussed as possible models for coronal X-ray sources, in particular
if the sources are separated into different components, including an
``above-the-loop-top'' source.

Here, we focus on the X1.3 April 25, 2014 partially occulted limb
flare, which shows a rich loop-top source morphology. Results on the
flare light curve in different energies, the source structure from
X-ray imaging, and a detailed spectral analysis of the different
source components from imaging spectroscopy are presented. We note
that this flare has also recently been studied in the context of CME
initiation and a magnetic breakout scenario. See the paper by Chen et
al. \cite{Chen-etal-2016} for further details on these
aspects.


\section{Temporal evolution}
\begin{figure}[h]
\centering
\includegraphics[width=0.63\textwidth]{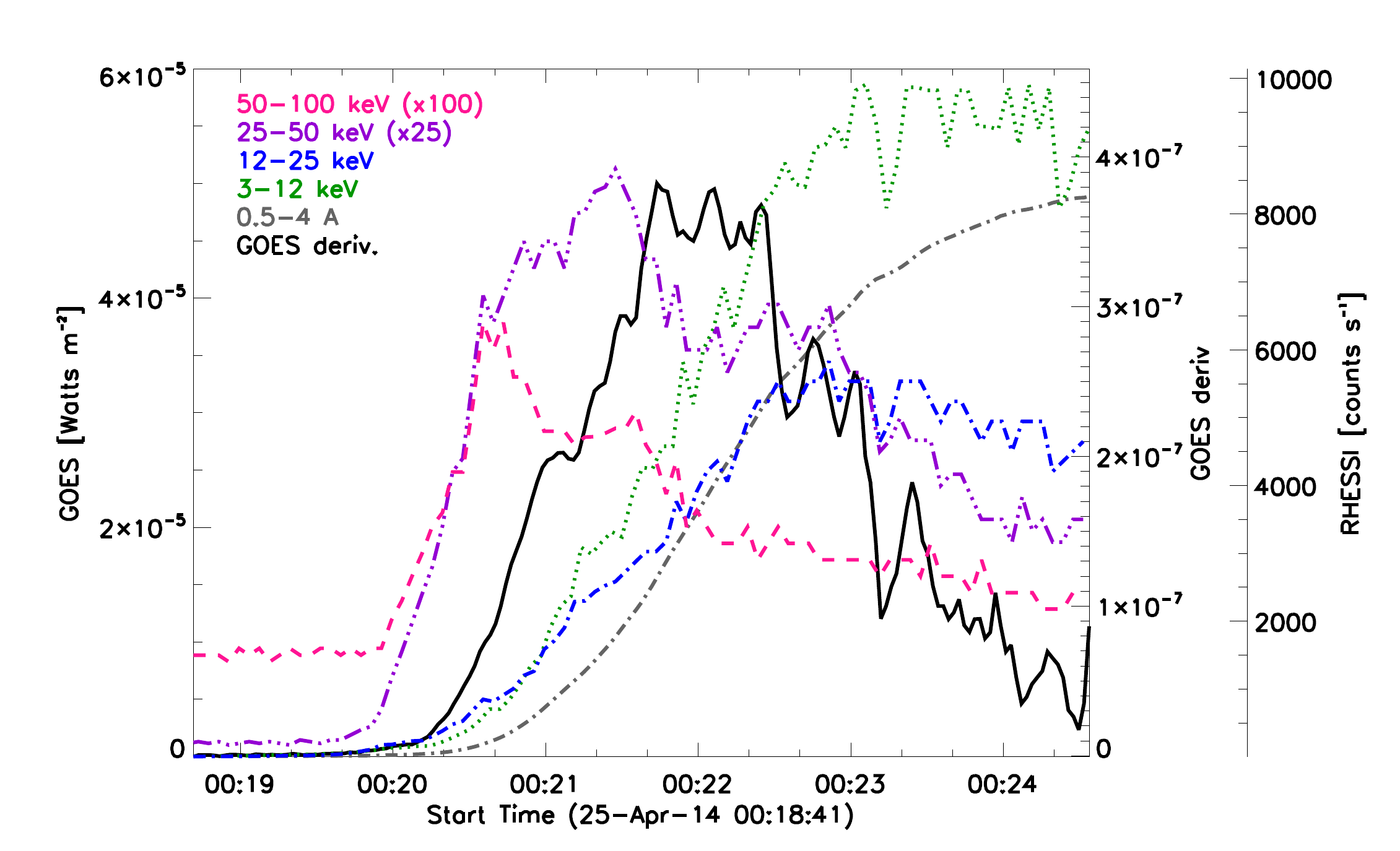}
\includegraphics[width=0.36\textwidth]{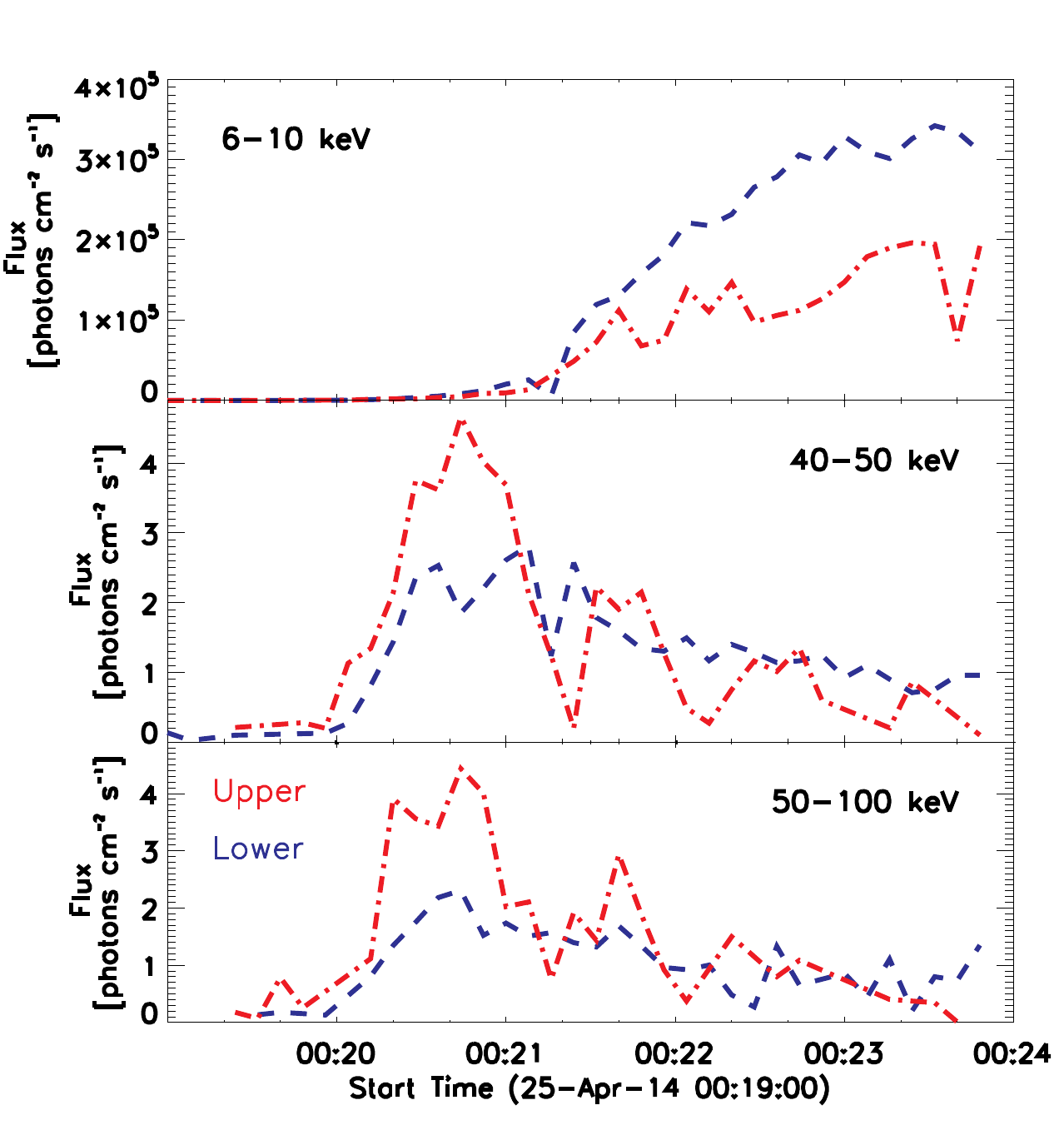}
\caption{\small Upper left panel: Light curves of \emph{RHESSI} count
  rates at four energy ranges (green, blue, purple and magenta),
  \emph{GOES} high energy flux (0.4 to 4 \AA, grey, dash-dotted) and
  its time derivative (black, solid). The counts in the two high
  energy channels are multiplied by 25 and 100, respectively, to make
  them comparable in magnitude. Right panel: Time evolution
  of photon flux in the two coronal regions selected (see the right
  panel of Figure~\ref{fig:images}), for three different energy
  channels.}
\label{fig:lightcurve}
\end{figure}
We analyzed the temporal evolution of the hard X-ray flux for this
flare and compared it with the \emph{GOES} soft X-ray evolution in the
high energy channel (0.4 to 4 \AA). The left panel of
Figure~\ref{fig:lightcurve} gives an overview of the \emph{RHESSI}
light curve in four energy channels and the correlation with the
\emph{GOES} flux and its time derivative.
   
It can be seen that the two lowest energy channels are delayed with
respect to the \emph{GOES} derivative and are roughly correlated with
the slow rise in \emph{GOES} flux. The two high energy channels
25-50~keV and 50-100~keV, on the other hand, have an earlier and
quicker rise to their maxima. In particular the 50-100~keV shows a
prominent peak at the early onset and then a fast decay. A
cross-correlation analysis showed the best correlation between the
25-50~keV channel and the \emph{GOES} derivative, with a correlation
coefficient of 0.81 and a lag of 12 s.

We note that a statistical study of more partially occulted
events is currently in progress and a significant number of flares show better
correlations and often small or zero lag. The implications of these
results, in particular in the context of the so-called Neupert effect
\cite{Neupert-1968,Veronig-etal-2005}, will be discussed elsewhere.

The right panel of Figure~\ref{fig:lightcurve} shows the photon flux
evolution in two different regions of the coronal source, namely an
upper and lower part. The regions are defined in the right panel of
Figure~\ref{fig:images}. We see that there is a faster rise at high
energies in the upper part of the coronal source resulting in a
stronger maximum. The lower energy fluxes slowly rise to larger values
for the lower source. We now turn to the details of the source
structure in X-ray images.

\section{Hard X-ray Imaging}
\begin{figure}[t]
\centering
\includegraphics[width=0.99\textwidth]{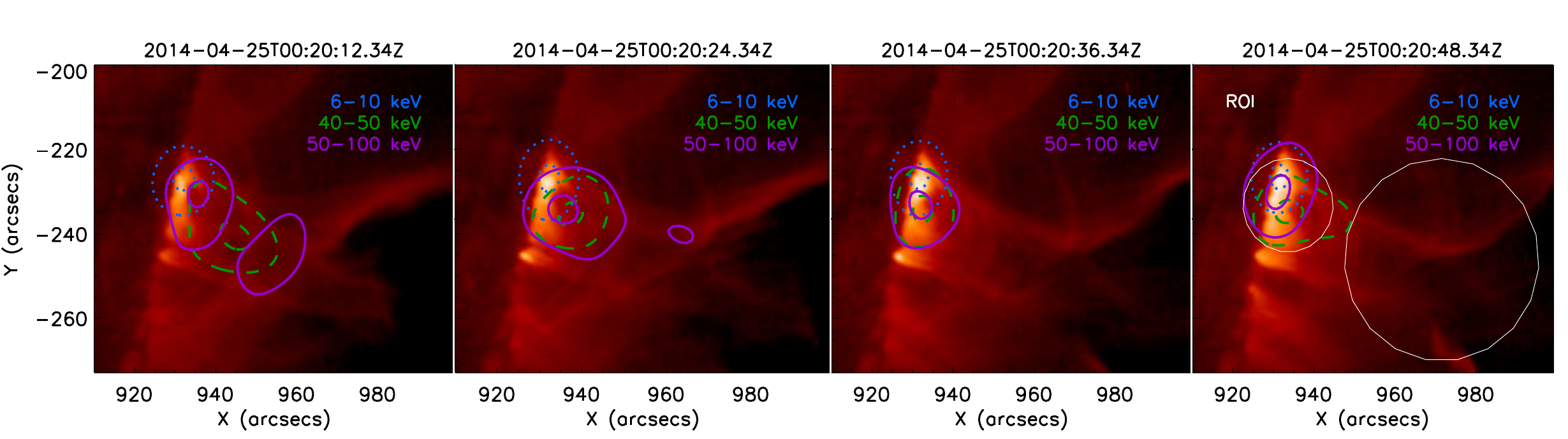}
\caption{\small Time evolution of the X1.3 April 25, 2014 flare in AIA
  131~\AA\ with \emph{RHESSI} contours at different energy ranges
  (77\% and 97\% of the maximum intensity at each time, applying the
  CLEAN algorithm): 6-10~keV (dotted blue), 40-50~keV (dashed green)
  and 50-100~keV (solid purple). The rightmost panel also indicates
  the two selected ROI for the imaging spectroscopy analysis of
  Section~\ref{sec:spectroscopy}.}
\label{fig:images}
\end{figure}
\emph{RHESSI}'s unique imaging capability allows a detailed study of
the spatial structure of the hard X-ray emission. In
Figure~\ref{fig:images}, we present the evolution over four subsequent
time intervals of 12 s length. The X-ray contours, obtained using the
CLEAN algorithm \cite{Hurford-etal-2002}, at three different energies
are overlaid over AIA 131~\AA\ background images closest in time to
the \emph{RHESSI} integration time interval (start times given above
the images). An expanding upper loop structure can be detected in the
AIA images, associated to a high energy (50-100 keV) upper source
moving outwards.

The high energy (up to 100 keV), non-thermal loop-top emission comes
mainly from the upper region of the X-ray source, while the thermal
6-10~keV emission is much stronger in the lower parts of the loops. We
note that no quasi-footpoint emission is detected, confirming that the
flare is partially occulted and that we are seeing a simultaneously
observed loop-top and a second higher above-the-loop-top source (see,
e.g., \cite{Liu-etal-2008,Liu-etal-2013,Oka-etal-2015}). These
uncommon observations allow us to separate the upper and lower coronal
emission for further studies.  To this end, we performed an imaging
spectroscopy study discussed next.

\section{Imaging spectroscopy}
\label{sec:spectroscopy}
Thanks to the high energy and spatial resolution of \emph{RHESSI}, it
is possible to obtain X-ray spectra of different separated regions. In
order to study the spectral evolution of the two sources, we defined
two regions of interest (ROI), indicated as white circles in the right
panel of Figure~\ref{fig:images}. We obtained photon spectra of the
two different regions resulting from the CLEAN images of these
regions.

\begin{figure}[t]
\centering
\includegraphics[width=0.99\textwidth]{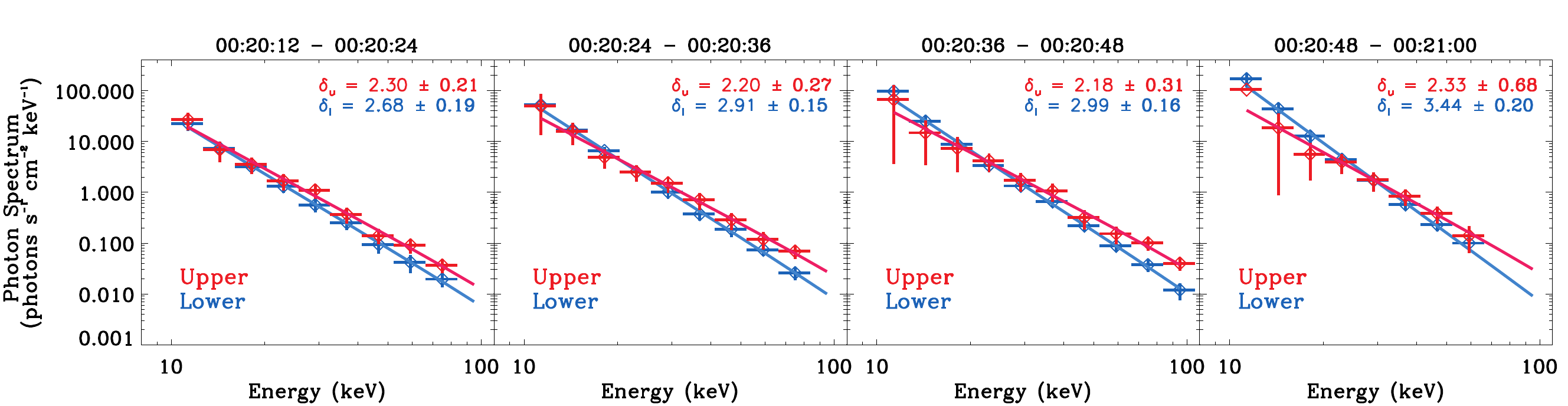}
\caption{\small Photon spectra at the same four time ranges as in
  Figure~\ref{fig:images} for the two separate ROI (red: upper source,
  blue: lower source). The spectra have been fitted to a thin-target
  model based on a broken power-law. The parameter values of the fit
  are given in the lower parts of the panels.}
\label{fig:IMspectra}
\end{figure}

The red and blue data points of Figure~\ref{fig:IMspectra} give the
results for the upper and lower sources, respectively, at four
different times. The spectra are fitted to a thin-target model with a
power law electron energy spectrum. The resulting electron spectral
index $\delta$ is reported in the bottom part of the figures. In
general, the upper source has a harder high energy spectrum.

The fit derived from the thin target model is in good agreement with
the data in all cases. We note that a thick-target model is often not
an appropriate model for coronal sources, due to the low column depth
of gas in the source (see, however, \cite{Fleishman-etal-2016}). In
particular, the relevant time scales of particle escape and energy
loss in the loop source are often such that particles tend to lose
only a fraction of their energy before they escape. This was found in
the study by Chen \& Petrosian \cite{Chen-Petrosian-2013} for an X3.9
and an M2.1 on disk flares, where these time scales could be determined
non-parametrically with an inversion procedure.

A thin-target kappa function fit \cite{Kasparova-Karlicky-2009} to the
photon spectra gives, as should be expected, almost identical values
for $\kappa$ and $\delta$. This is due to the fact that $\kappa$ is
defined to be the parameter in the number density distribution while
$\delta$ is the index of the electron flux density distribution, see
the discussion in \cite{Oka-etal-2013}. We do not show these fit
results since they look very similar to Figure~\ref{fig:IMspectra}.

\begin{figure}[t]
\centering
\includegraphics[width=0.99\textwidth]{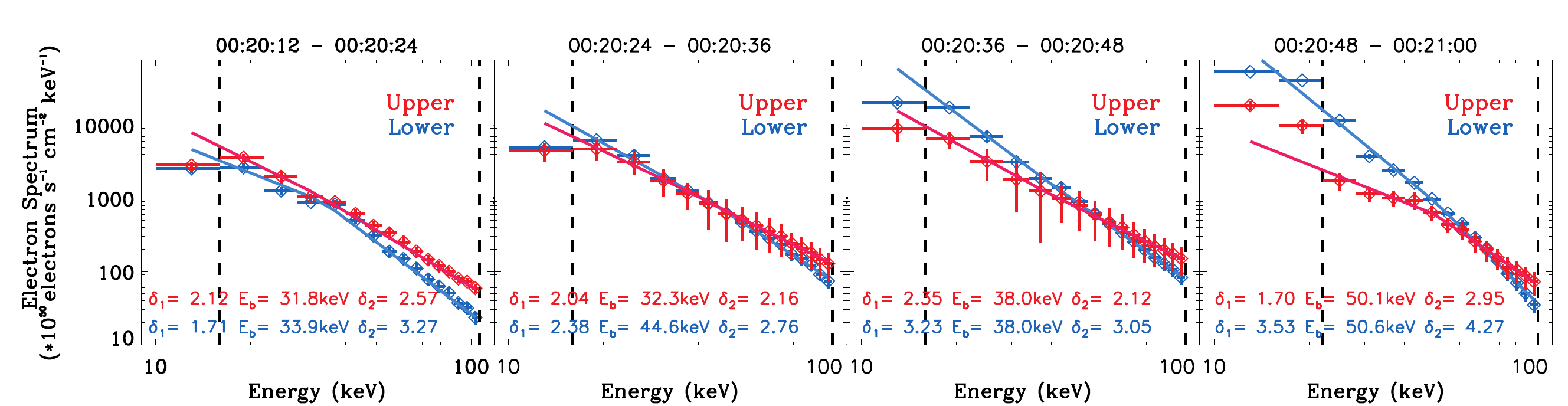}
\caption{\small Similar to Figure~\ref{fig:IMspectra}, but now as
  electron spectra resulting from the regularized imaging inversion
  with the same ROI. The spectra have been fitted to a broken
  power-law, with the parameter values of the fit reported in the
  lower parts of the panels.}
\label{fig:IMElectron_spectra}
\end{figure}

To gain an additional perspective on the spectral properties of
electrons, we obtained electron images and spectra directly using a
regularized imaging inversion method \cite{Piana-etal-2007}. The
resulting electron images (not shown) display a similar upward motion
of the upper source as in the photon images.  The electron spectra for
the same ROI are shown in Figure~\ref{fig:IMElectron_spectra}, and
were fitted to a broken power-law. We see that there is significant
electron flux at high energies, with clear power-laws extending to up
to ${\sim}100$~keV, for both source regions. Considering that the
errors in the fitting parameters are of the same order as in the
previous thin-target fit, the upper source has generally slightly
harder spectral indices and higher fluxes at high energies, pointing
at a possibly more effective acceleration mechanism in this part of
the loop. Most of the spectra are compatible
with a single power-law, in particular for the two time intervals in
the middle.

Taking into account the confidence intervals for the photon spectra
thin target fitting parameters, as stated in
Figure~\ref{fig:IMspectra}, we find that the resulting electron
spectra are in reasonable agreement for both methods. This confirms
the notion of thin-target emission for the entire coronal source
region and gives independent support to the different spectral
characteristics of the two parts of the source.

\section{Conclusions}
We have analyzed the X1.3 April 25, 2014 partially occulted limb flare
in hard X-rays with \emph{RHESSI} imaging and spectroscopy. The
coronal emission shows a rich morphology with at least two different
coronal source regions. The upper source is moving upwards and
detaching from the lower source during the flare evolution over a
short time interval of tens of seconds (see also
\cite{Sui-Holman-2003,Sui-etal-2004} for similar observations of other
flares).

We found a different electron spectral behavior in the upper and lower
source derived from a forward thin-target fit and regularized imaging
inversion. The differences between the results obtained with the two
methods are mostly within the error bounds. The upper source has
somewhat smaller (harder) spectral indices than the lower
source.\ Different scenarios could explain this result. Assuming the
reconnection region lies between these two source regions, the
effective distance to the reconnection region could imply different
turbulence levels. This in turn can result in harder (higher
turbulence level) or softer (lower turbulence level) electron spectra
for the accelerated population
\cite{Liu-etal-2008,Petrosian-Liu-2004}.

For further insights into the differences of coronal source modeling
and electron properties, a larger sample of partially occulted flares
needs to be studied. Detailed comparisons with simulations of particle
acceleration and transport will help to uncover the acceleration
mechanisms at work in different parts of the flare loop.
 
\small \ack
The authors thank S. Krucker for suggesting the combined \emph{RHESSI}
and \emph{GOES} light-curve analysis. F.E. thanks the organizers and
participants of the 15th Annual International Astrophysics Conference
for the stimulating discussions. This work was supported by NASA
grants NNX14AG03G and NNX13AF79G.

\section*{References}
\hspace{0.5cm}
\bibliographystyle{iopart-num}
\bibliography{references}

\providecommand{\newblock}{}
\begin{thebibliography}{10}
\expandafter\ifx\csname url\endcsname\relax
  \def\url#1{{\tt #1}}\fi
\expandafter\ifx\csname urlprefix\endcsname\relax\def\urlprefix{URL }\fi
\providecommand{\eprint}[2][]{\url{#2}}

\bibitem{Fletcher-etal-2011}
{Fletcher} L, {Dennis} B~R, {Hudson} H~S {\em et~al.\/} 2011 {\em \ssr\/} {\bf
  159} 19--106

\bibitem{Frost-Dennis-1971}
{Frost} K~J and {Dennis} B~R 1971 {\em \apj\/} {\bf 165} 655

\bibitem{Roy-Datlowe-1975}
{Roy} J~R and {Datlowe} D~W 1975 {\em \solphys\/} {\bf 40} 165--182

\bibitem{McKenzie-1975}
{McKenzie} D~L 1975 {\em \solphys\/} {\bf 40} 183--191

\bibitem{Hudson-1978}
{Hudson} H~S 1978 {\em \apj\/} {\bf 224} 235--240

\bibitem{Masuda-etal-1994}
{Masuda} S, {Kosugi} T, {Hara} H, {Tsuneta} S and {Ogawara} Y 1994 {\em \nat\/}
  {\bf 371} 495--497

\bibitem{Petrosian-etal-2002}
{Petrosian} V, {Donaghy} T~Q and {McTiernan} J~M 2002 {\em \apj\/} {\bf 569}
  459--473

\bibitem{Krucker-etal-2007}
{Krucker} S, {White} S~M and {Lin} R~P 2007 {\em \apjl\/} {\bf 669} L49--L52

\bibitem{Simoes-Kontar-2013}
{Sim{\~o}es} P~J~A and {Kontar} E~P 2013 {\em \aap\/} {\bf 551} A135

\bibitem{Petrosian-2012}
{Petrosian} V 2012 {\em \ssr\/} {\bf 173} 535--556

\bibitem{Lin-etal-2002}
{Lin} R~P, {Dennis} B~R, {Hurford} G~J {\em et~al.\/} 2002 {\em \solphys\/}
  {\bf 210} 3--32

\bibitem{Krucker-Lin-2008}
{Krucker} S and {Lin} R~P 2008 {\em \apj\/} {\bf 673} 1181--1187

\bibitem{Kasparova-Karlicky-2009}
{Ka{\v s}parov{\'a}} J and {Karlick{\'y}} M 2009 {\em \aap\/} {\bf 497}
  L13--L16

\bibitem{Oka-etal-2013}
{Oka} M, {Ishikawa} S, {Saint-Hilaire} P, {Krucker} S and {Lin} R~P 2013 {\em
  \apj\/} {\bf 764} 6

\bibitem{Bian-etal-2014}
{Bian} N~H, {Emslie} A~G, {Stackhouse} D~J and {Kontar} E~P 2014 {\em \apj\/}
  {\bf 796} 142

\bibitem{Chen-etal-2016}
{Chen} Y, {Du} G, {Zhao} D, {Wu} Z, {Liu} W, {Wang} B, {Ruan} G, {Feng} S and
  {Song} H 2016 {\em \apjl\/} {\bf 820} L37

\bibitem{Neupert-1968}
{Neupert} W~M 1968 {\em \apjl\/} {\bf 153} L59

\bibitem{Veronig-etal-2005}
{Veronig} A~M, {Brown} J~C, {Dennis} B~R, {Schwartz} R~A, {Sui} L and {Tolbert}
  A~K 2005 {\em \apj\/} {\bf 621} 482--497

\bibitem{Hurford-etal-2002}
{Hurford} G~J, {Schmahl} E~J, {Schwartz} R~A {\em et~al.\/} 2002 {\em
  \solphys\/} {\bf 210} 61--86

\bibitem{Liu-etal-2008}
{Liu} W, {Petrosian} V, {Dennis} B~R and {Jiang} Y~W 2008 {\em \apj\/} {\bf
  676} 704-716

\bibitem{Liu-etal-2013}
{Liu} W, {Chen} Q and {Petrosian} V 2013 {\em \apj\/} {\bf 767} 168

\bibitem{Oka-etal-2015}
{Oka} M, {Krucker} S, {Hudson} H~S and {Saint-Hilaire} P 2015 {\em \apj\/} {\bf
  799} 129

\bibitem{Fleishman-etal-2016}
{Fleishman} G~D, {Xu} Y, {Nita} G~N and {Gary} D~E 2016 {\em \apj\/} {\bf 816}
  62

\bibitem{Chen-Petrosian-2013}
{Chen} Q and {Petrosian} V 2013 {\em \apj\/} {\bf 777} 33 (\textit{Preprint}
  \eprint{1307.1837})

\bibitem{Piana-etal-2007}
{Piana} M, {Massone} A~M, {Hurford} G~J, {Prato} M, {Emslie} A~G, {Kontar} E~P
  and {Schwartz} R~A 2007 {\em \apj\/} {\bf 665} 846--855

\bibitem{Sui-Holman-2003}
{Sui} L and {Holman} G~D 2003 {\em \apjl\/} {\bf 596} L251--L254

\bibitem{Sui-etal-2004}
{Sui} L, {Holman} G~D and {Dennis} B~R 2004 {\em \apj\/} {\bf 612} 546--556

\bibitem{Petrosian-Liu-2004}
{Petrosian} V and {Liu} S 2004 {\em \apj\/} {\bf 610} 550--571

\end{thebibliography}

\end{document}